\documentclass[pra, twocolumn, showpacs, superscriptaddress]{revtex4-1}

\usepackage{graphicx,amsmath,amssymb}

\begin{document}

\title{Route to observing topological edge modes in ultracold fermions}
\author{Junjun Xu}
\affiliation{Department of Physics, University of Science and Technology Beijing, Beijing 100083, China}
\affiliation{Laboratory of Atomic and Solid State Physics, Cornell University, Ithaca, New York 14853, USA}
\author{Qiang Gu}
\affiliation{Department of Physics, University of Science and Technology Beijing, Beijing 100083, China}
\author{Erich J. Mueller}
\affiliation{Laboratory of Atomic and Solid State Physics, Cornell University, Ithaca, New York 14853, USA}
\date{\today}

\begin{abstract}
We show how to exploit the rich hyperfine structure of fermionic alkali-metal atoms to produce a quasi-1D topological superfluid while avoiding excessive heating from off-resonant scattering. We model interacting fermions where four hyperfine states are coupled by a variety of optical and microwave fields. We calculate the local density of states in a trap, finding regimes with zero energy topological edge modes. Heating rates in this system are significantly suppressed compared to simple Raman-induced spin-orbit coupling approaches. We also estimate the two- and three-body decay rates and find a reasonable lifetime at small, but experimentally relevant densities.
\end{abstract}
 
\maketitle

\section{Introduction}
Experimental demonstrations of Raman-induced spin-orbit coupling in cold atoms have motivated theoretical proposals for using these techniques to observe topological edge modes \cite{Lin, Cheuk, Wang, Mueller, Zhai, Galitski, Wei1, Liu1, Liu2, Liu3}. A practical concern with these proposals is that the Raman process introduces an intrinsic heating mechanism. As discussed by Wei and Mueller \cite{Wei2}, this heating is particularly problematic for $^6{\rm Li}$, one of the work-horse fermions used in cold atom experiments. Here we show how to exploit the rich hyperfine structure of alkali atoms to ameliorate this heating.

Excitement about Raman-induced spin-orbit coupling in cold atoms is related to broader interest in ``topological" states of matter, i.e., those which are characterized by topological invariants which lead to non-trivial edge modes. Spin-orbit coupling is an essential ingredient in many electronic topological states \cite{Kane, Fu, Hasan, Qi}, and these Raman experiments promise to enable cold-atom studies of analogous physics. A particularly natural setup would involve a gas of fermionic atoms in a quasi-1D ``wire-trap''. A magnetic-field induced Feshbach resonance could drive the atoms into a paired superfluid state. Adding spin-orbit coupling would lead to Majorana edge modes \cite{Wei1, Liu1, Liu2, Liu3, Oreg}.

The same photons which provide spin-orbit coupling also heat the cloud. Here we propose minimizing this heating by dividing our atoms into two populations, distinguished by their hyperfine spins. Both the Feshbach resonant interactions and the Raman couplings are state selective, and one of these populations can be chosen to be strongly interacting, yet feel no spin-orbit coupling. The other will be essentially non-interacting, but experience a substantial spin-orbit coupling. The Raman lasers will produce heat, but the heating rate will be proportional to the number of atoms in the Raman-coupled states. By making the superfluid population large compared to the Raman-coupled population, one can make studying this system practical.

\begin{figure}[th]
  \includegraphics[width=0.47\textwidth]{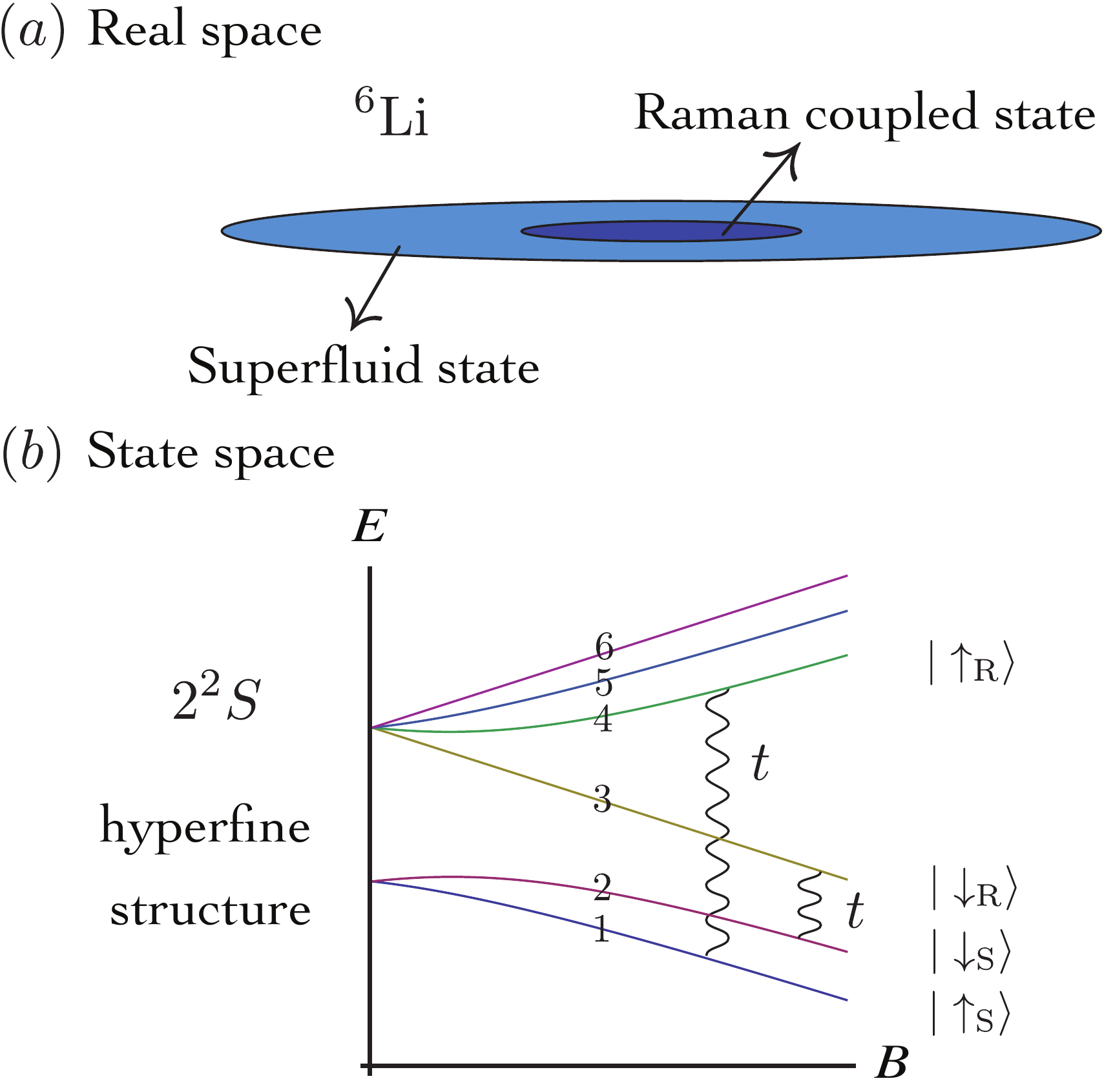}
  \caption{(Color online) Illustration of our proposal in case of $^6{\rm Li}$. (a) Schematic of proposed experiment in real space. There is a large superfluid reservoir composed of $\vert\uparrow_{\rm S}\rangle$ and $\vert\downarrow_{\rm S}\rangle$ fermions overlapping a smaller cloud of Raman-induced spin-orbit coupled particles with spin states $\vert\uparrow_{\rm R}\rangle$ and $\vert\downarrow_{\rm R}\rangle$. (b) Hyperfine structure of $^6{\rm Li}$, illustrating the states used in the proposed experiment. The tunneling $t$ is turned on by micro (radio) waves between $\vert\uparrow_{\rm S}\rangle$ ($\vert\downarrow_{\rm S}\rangle$) and $\vert\uparrow_{\rm R}\rangle$ ($\vert\downarrow_{\rm R}\rangle$).}
  \label{fig:fig1}
\end{figure}

\begin{figure}[th]
  \includegraphics[width=0.4\textwidth]{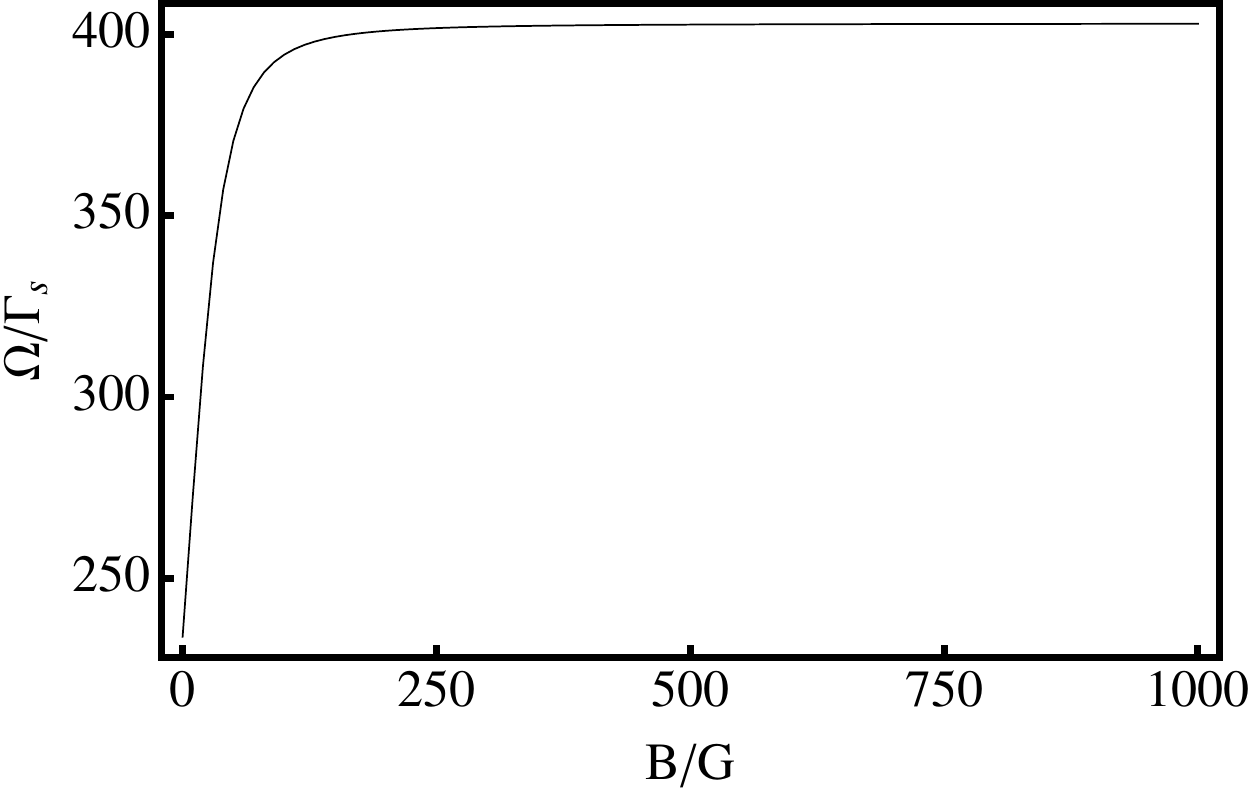}
  \caption{The ratio of Raman coupling strength $\Omega$ to inelastic scattering rate $\Gamma_s$ as a function of magnetic field for transition between the third and fourth lowest hyperfine states of $^6{\rm Li}$ \cite{Wei3}.}
  \label{fig:fig2}
\end{figure}

Figure \ref{fig:fig1} shows the hyperfine level diagram of $^6{\rm Li}$, and the states we propose using for this experiment. The strongly interacting population ($\vert\uparrow_{\rm S}\rangle$ and $\vert\downarrow_{\rm S}\rangle$) will be formed from the lowest and second lowest hyperfine states. These states have a Feshbach resonance at $B=832{\rm G}$ in 3D \cite{Chin, Zurn}. The superfluid formed by these states has been observed \cite{Zwierlein}. The Raman-coupled population ($\vert\downarrow_{\rm R}\rangle$ and $\vert\uparrow_{\rm R}\rangle$) will be formed from the third and fourth lowest hyperfine states. Figure \ref{fig:fig2} shows the ratio of Raman coupling strength to inelastic scattering rate as a function of magnetic field for $\vert\uparrow_{\rm R}\rangle$ and $\vert\downarrow_{\rm R}\rangle$ states. At fields near the Feshbach resonance, these matrix elements are comparable to those used in prior Raman-induced spin-orbit coupling experiments. Interconversion between $\vert\downarrow_{\rm S}\rangle$ and $\vert\downarrow_{\rm R}\rangle$ can be driven by Radio waves. Microwaves can independently drive the $\vert\uparrow_{\rm S}\rangle$ to $\vert\uparrow_{\rm R}\rangle$ transition. By appropriately detuning these transitions, one can set the relative equilibrium populations of each of the states. In the presence of such interconversion, we find that there are parameters for which ground state will possess Majorana edge modes.

In addition to reducing heating, our two-population approach avoids the difficult problem of finding spin states which can both have strong interaction and be coupled by Raman lasers. The R-atoms have no readily accessible Feshbach resonances, and the Raman coupling strengths for the S-atoms are 40 times smaller than those of the R-atoms (when the intensity is adjusted to give the same inelastic scattering rates). Furthermore, our approach provides a nice analogy to the solid state experiments, where the superconductivity and the spin-orbit coupling comes from separate materials.

It is natural to ask if similar physics could be seen in higher dimension. Unfortunately, current cold atom techniques give a unidirectional spin-orbit coupling, and do not generalize to 2D or 3D.

In the remainder of this paper, we model our proposed experiment. In section II we write a Hamiltonian for the system, and explain how it can be analyzed. In section III we show a Bogoliubov-de Gennes calculation of the properties of this system. We calculate the local density of states, which can be studied with spatially resolved radio frequency spectroscopy.  We find two regimes, one of which possesses topological edge modes. We further explain this physics through a local density approximation. At the end, we estimate the heating and collisional losses in our system.

\section{Theoretical Model}
We model our system with a Hamiltonian
\begin{eqnarray}
H=H_{\rm SF}+H_{\rm RM}+H_{\rm T},
\end{eqnarray}
where 
\begin{eqnarray}
H_{\rm SF}=&&\sum_{\sigma=\uparrow, \downarrow}\int dx\psi_{\sigma}^{\dagger}(x)\left(-\frac{\hbar^2\nabla^2}{2m}-\mu_{\rm S}\right)\psi_{\sigma}(x)\nonumber\\
&&+\int dx\left(\Delta\psi_{\uparrow}^\dagger(x)\psi_{\downarrow}^\dagger(x)+{\rm H.c.}\right)
\end{eqnarray}
describes the energy of the strongly interacting reservoir within a mean field approximation. The field operator $\psi_{\sigma}(x)$ corresponds to atoms with spin $\sigma=\uparrow, \downarrow$. Here $\mu_{\rm S}=\mu_{\rm SF}-V(x)$ with $\mu_{\rm SF}$ the chemical potential of these atoms, and $V(x)=m\omega^2x^2/2$ is the harmonic potential with frequency $\omega$. The superfluid energy gap $\Delta$ is taken to be real and positive. The Hamiltonian of the Raman-coupled atoms is
\begin{eqnarray}
H_{\rm RM}=&&\sum_{\sigma=\uparrow, \downarrow}\int dx\varphi_{\sigma}^{\dagger}(x)\left(-\frac{\hbar^2\nabla^2}{2m}-\mu_{\rm R}\right)\varphi_{\sigma}(x)\nonumber\\
&&+\int dx\left(\lambda(x)\varphi_{\uparrow}^{\dagger}(x)\varphi_{\downarrow}(x)+{\rm H.c.}\right),\end{eqnarray}
with field operator $\varphi_{\sigma}(x)$, and $\mu_{\rm R}=\mu_{\rm SF}-V(x)-\delta$ where $\delta$ is the RF detuning. The Raman coupling is described by $\lambda(x)=\Omega e^{2ik_rx}$ where $\Omega$ is the two-photon Rabi rate and $k_r$ is the recoil momentum. The radio and microwave driven tunneling between the superfluid and Raman states are characterized by
\begin{eqnarray}
H_{\rm T}=t\sum_{\sigma=\uparrow, \downarrow}\int dx\left(\varphi_{\sigma}^{\dagger}(x)\psi_{\sigma}(x)+{\rm H.c.}\right),
\end{eqnarray}
with tunneling strength $t$, proportional to the intensity of these fields. 

The full Hamiltonian can be written in a matrix form
\begin{eqnarray}
H=&&\int dx \left[\Psi_{\rm S}^\dagger, \Psi_{\rm R}^\dagger\right]
\left[\begin{array}{cc}
\mathcal{H}_{\rm S} & \mathcal{H}_{\rm T} \\
\mathcal{H}_{\rm T} & \mathcal{H}_{\rm R}
\end{array}\right]
\left[\begin{array}{c}
\Psi_{\rm S}\\
\Psi_{\rm R}
\end{array}\right],\\
=&&\int dx \Psi^\dagger\mathcal{H}\Psi
\end{eqnarray}
where $\Psi_{\rm S}=\left[\psi_\uparrow(x), \psi_\downarrow(x), \psi_\uparrow^\dagger(x), \psi_\downarrow^\dagger(x)\right]^T$ and $\Psi_{\rm R}=\left[\varphi_\uparrow(x), \varphi_\downarrow(x), \varphi_\uparrow^\dagger(x),\varphi_\downarrow^\dagger(x)\right]^T$ are length 4 vectors of field operators, and the equality defines $\Psi$ and $\mathcal{H}$. The matrix Hamiltonian density is made of
\begin{equation}
\mathcal{H_{\rm S}}=
\left[\begin{array}{cccc}
    \mathcal{H}_0-\mu_{\rm S} & 0 & 0 & \Delta  \\ 
    0 & \mathcal{H}_0-\mu_{\rm S} & -\Delta & 0  \\ 
    0 & -\Delta & -\mathcal{H}_0+\mu_{\rm S} & 0 \\ 
    \Delta & 0 & 0 & -\mathcal{H}_0+\mu_{\rm S} \\   
\end{array}\right],\nonumber
\end{equation}
\begin{equation}
\mathcal{H_{\rm R}}=
\left[\begin{array}{cccc}
    \mathcal{H}_0-\mu_{\rm R} & \lambda(x) & 0 & 0  \\ 
    \lambda^*(x) & \mathcal{H}_0-\mu_{\rm R} & 0 & 0  \\ 
    0 & 0 & -\mathcal{H}_0+\mu_{\rm R} & -\lambda^*(x) \\ 
    0 & 0 & -\lambda(x) & -\mathcal{H}_0+\mu_{\rm R} \\   
\end{array}\right],\nonumber
\end{equation}
\begin{equation}
\mathcal{H}_{\rm T}=
\left[\begin{array}{cccc}
    t & 0 & 0 & 0  \\ 
    0 & t & 0 & 0  \\ 
    0 & 0 & -t & 0 \\ 
    0 & 0 & 0 & -t \\   
\end{array}\right].\nonumber
\end{equation}
Here $\mathcal{H}_0=-\hbar^2\nabla^2/(2m)$. We expand the field operator in a complete basis
\begin{eqnarray}
\hat{\Psi}_i(x)=\sum_nc_{ni}(x)\hat{\gamma}_n.
\label{eq:psi}
\end{eqnarray}
For clarity, in Eq. (\ref{eq:psi}) we have added hats to all operators and explicitly show all $x$ dependence. (For notational simplicity, these typographic clues are left out at our other equations.) The index $i$ runs from 1 to 8, and $n$ is summed over all eigenstates. Here $\gamma_n$ is the excitation operator with energy $E_n$:
\begin{eqnarray}
H=1/2\sum_nE_n\gamma^\dagger_n\gamma_n+{\rm const}.
\label{eq:eng}
\end{eqnarray}
The energies are found from the eigenvalue problem,
\begin{eqnarray}
\mathcal{H}c_n=E_nc_n,
\label{eq:bdg}
\end{eqnarray}
where $c_n(x)$ is a 8$\times$1 matrix with the $i$-th row element $c_{ni}(x)$. In this particle-hole symmetric representation of the system, we have $\gamma_n=\gamma_{-n}^\dagger$, and each mode effectively appears twice in Eq. (\ref{eq:eng}). The Majorana states corresponds to the zero energy states, i.e., $\gamma_0=\gamma_{0}^\dagger$.

We numerically solve the Bogoliubov-de Gennes equation, Eq. (\ref{eq:bdg}). The local density of states of superfluid and Raman coupled particles then can be written as 
\begin{eqnarray}
D_{\uparrow(\downarrow)\rm{S}}(x, E)=\sum_n\vert c_{n1(2)}(x)\vert^2\delta(E-E_n),\\
D_{\uparrow(\downarrow)\rm{R}}(x, E)=\sum_n\vert c_{n5(6)}(x)\vert^2\delta(E-E_n),
\end{eqnarray}
where the summation is over all eigenstates. 

\begin{figure}[h]
  \includegraphics[width=0.48\textwidth]{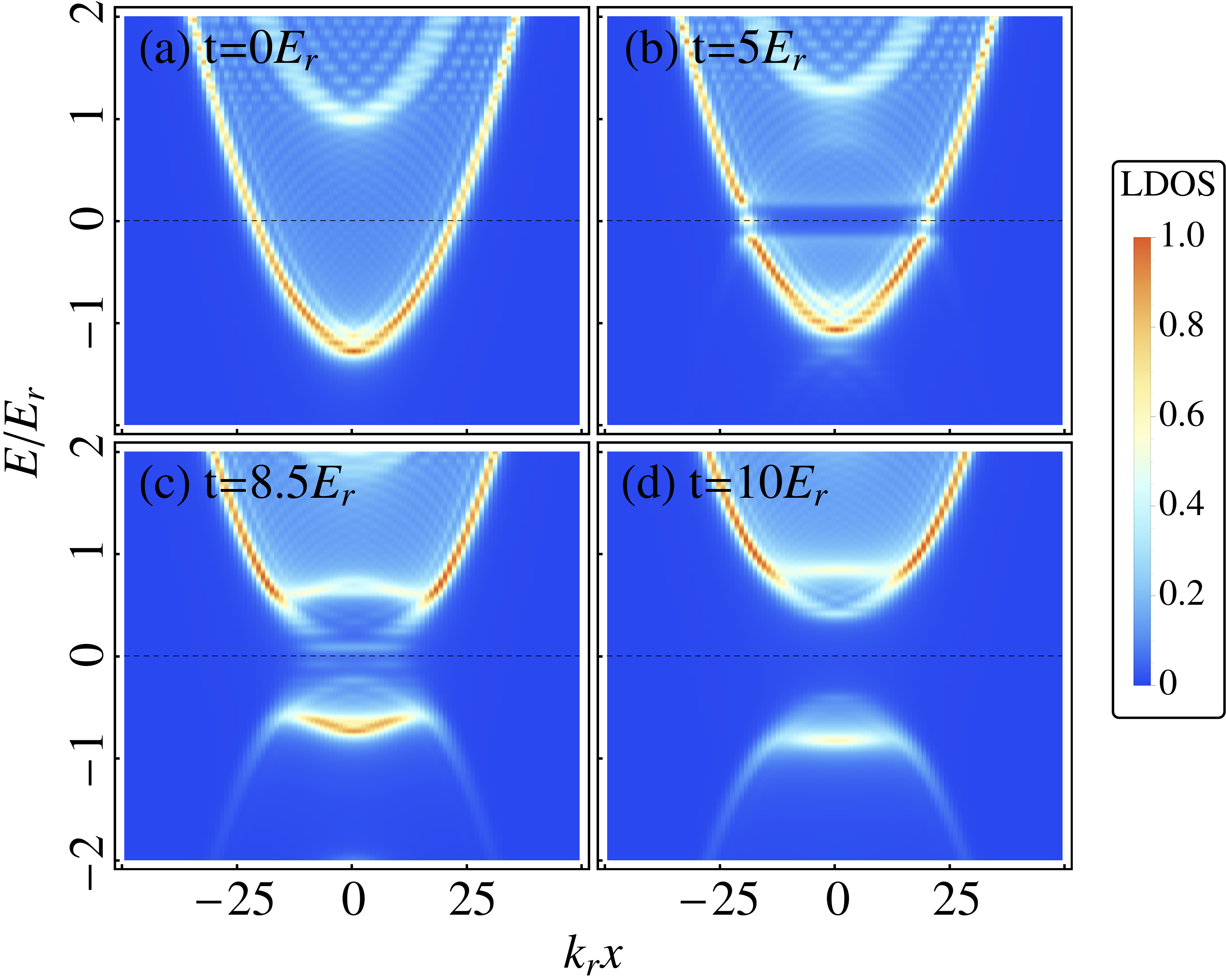}
  \caption{(Color online) Local density of states (LDOS) for the Raman-coupled atoms with different tunneling strength $t$. Energies $E$ are measured in terms of the recoil energy, $E_r=\hbar^2k_r^2/(2m)$ where $k_r$ is the wave-vector of the Raman lasers. In this calculation we choose $\Delta=50E_r$, $\Omega=E_r$. The chemical potential is set to be $\mu_{\rm SF}=50E_r$ and $\delta=49E_r$. The harmonic potential $V(x)=m\omega^2x^2/2$ has $\hbar\omega=0.1E_r$. The horizontal dashed lines show $E=0$. (a) Topologically trivial state without tunneling. (b) Topologically non-trivial state when zero energy modes appear on the edges. (c) Transition from non-trivial to trivial state when edge modes begin to disappear. (d) Topologically trivial state at large $t$.}
  \label{fig:fig3}
\end{figure}

\section{Topological edge modes}

Our results are summarized by the local density of states shown in Fig. \ref{fig:fig3}. In Fig. \ref{fig:fig3} (b) we clearly see zero energy modes near the edge of the cloud. These are Majorana modes. At stronger coupling $t$, this feature disappears. Below we will present an argument for when Majorana modes will be present.

\begin{figure}[th]
  \includegraphics[width=0.4\textwidth]{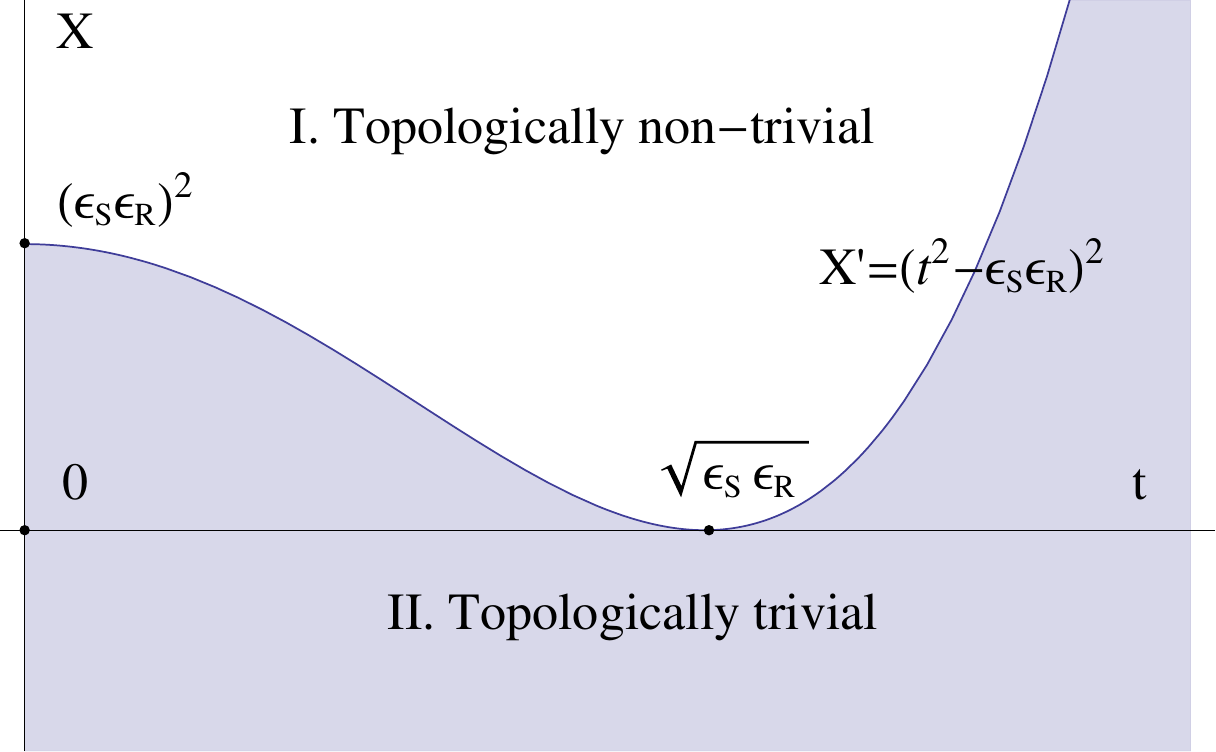}
  \caption{(Color online) Illustration of phase diagram of tunneling induced superfluid-Raman coupled system as a function of $t$ and $X\equiv\Omega^2\left(\Delta^2+\epsilon_{\rm S}^2\right)-\Delta^2\epsilon_{\rm R}^2$. The diagram can be separated into two phases: I. Topologically non-trivial state which has zero energy modes at the edge and II. Topologically trivial state where no Majorana modes should be found. The curve shows where $X=\left(t^2-\epsilon_{\rm S}\epsilon_{\rm R}\right)^2\equiv X^\prime$.}
  \label{fig:fig4}
\end{figure}

First we rewrite the Hamiltonian density in momentum space
\begin{equation}
\mathcal{H}_k=
\left[\begin{array}{cccccccc}
 \epsilon_{k{\rm S}} & 0 & 0 & \Delta & t & 0 & 0 & 0\\
 0 & \epsilon_{k{\rm S}} & -\Delta & 0 & 0 & t & 0 & 0\\
 0 & -\Delta & -\epsilon_{k{\rm S}} & 0 & 0 & 0 & -t & 0\\
 \Delta & 0 & 0 & -\epsilon_{k{\rm S}} & 0 & 0 & 0 & -t\\
 t & 0 & 0 & 0 & \epsilon_{k{\rm R}} & \Omega & 0 & 0\\
 0 & t & 0 & 0 & \Omega & \epsilon_{-k{\rm R}} & 0 & 0\\
 0 & 0 & -t & 0 & 0 & 0 & -\epsilon_{k{\rm R}} & -\Omega\\
 0 & 0 & 0 & -t & 0 & 0 & -\Omega & -\epsilon_{-k{\rm R}}
\end{array}\right],
\label{eq:hk}
\end{equation}
where $\epsilon_{k{\rm S}}=\hbar^2(k+k_r)^2/2m-\mu_{\rm S}$, $\epsilon_{k{\rm R}}=\hbar^2(k+k_r)^2/2m-\mu_{\rm R}$, and $\epsilon_{-k{\rm R}}=\hbar^2(k-k_r)^2/2m-\mu_{\rm R}$. The eigenstates of Eq. (\ref{eq:hk}) define a mapping from $U(1)$ [the values of $k$] to $C_8$ [8-dimensional complex space]. As explicitly explored in \cite{Wei1}, these mappings can be divided into topological distinct equivalence classes. As one varies the parameters $\Delta$, $\mu$, $t$, etc., the topological classification of the bands do not change, unless two bands touch. The locations of the Majorana modes in the trap are roughly the points where the spatially changing $\mu(r)$ leads to exactly such a band crossing. By symmetry, we expect the band crossings to occur at $k=0$, and they can be found by setting ${\rm det}(\mathcal{H}_{k=0})=0$. This yields the simple condition
\begin{eqnarray}
\left(t^2-\epsilon_{\rm S}\epsilon_{\rm R}\right)^2=\Omega^2\left(\Delta^2+\epsilon_{\rm S}^2\right)-\Delta^2\epsilon_{\rm R}^2.
\label{eq:phase}
\end{eqnarray}
where $\epsilon_{\rm S}=\hbar^2k_r^2/2m-\mu_{\rm S}$ and $\epsilon_{\rm R}=\hbar^2k_r^2/2m-\mu_{\rm S}+\delta$. Figure \ref{fig:fig4} shows the phase diagram of the homogeneous system, resulting from this argument. Further treating $\mu_{\rm S}=\mu_{\rm SF}$ as the chemical potential at the center of the trap, there will be a Majorana mode somewhere in the cloud if the system is locally topologically non-trivial at the center. The plot of $X^\prime=\left(t^2-\epsilon_{\rm S}\epsilon_{\rm R}\right)^2$ in Fig. \ref{fig:fig4} separates the phase diagram into two phases: I. Topologically non-trivial phase and II. Topologically trivial phase. The boundary is a non-monotonic function of $t$, however at large enough $t$, the state is always topologically trivial. The line $t=0$ also corresponds to a band crossing, but if one reflects Fig. \ref{fig:fig4} across the vertical axis, the picture remain the same.

\begin{figure}[th]
  \includegraphics[width=0.38\textwidth]{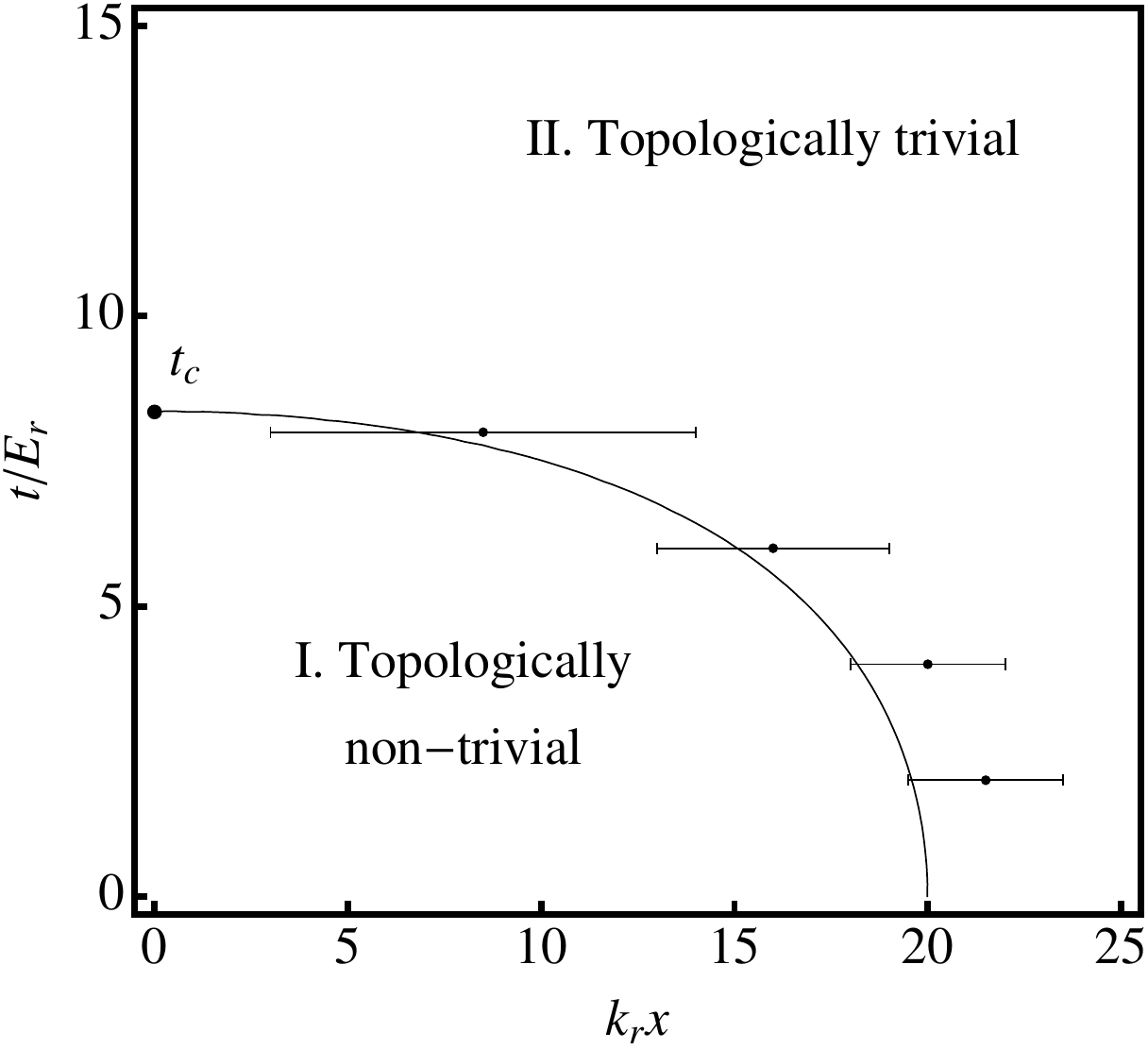}
  \caption{Local topology in the trap. $E_r=\hbar^2k_r^2/(2m)$ is the recoil energy with $k_r$ the recoil momentum from the Raman laser, and $t$ is the strength of coupling between the superfluid and Raman coupled states (see Fig. \ref{fig:fig1}). Parameters are chosen as in Fig. \ref{fig:fig3}. The solution of Eq. (\ref{eq:phase}) (the curve) separates the diagram into two regimes: I. Locally topologically non-trivial phase and II. Locally topologically trivial phase. The edge of the trapped cloud is always trivial. Within a local density approximation (LDA) Majorana edge modes appear at the spatial boundary between these phases, and will be found if $t\leqslant t_c\approx 8.4E_r$. The points with ``error bars" show the location and width of the Majorana mode in numerical simulations using the parameters in Fig. \ref{fig:fig3}. The discrepancy between the numerical calculations and LDA becomes smaller as the number of particles is increased.}
  \label{fig:fig5}
\end{figure}

Treating the $\mu_{\rm S}$ and $\mu_{\rm R}$ in Eq. (\ref{eq:phase}) as spatially dependent: $\mu_{\rm S}=\mu_{\rm SF}-V(x)$, $\mu_{\rm R}=\mu_{\rm SF}-V(x)-\delta$, the relationship in Eq. (\ref{eq:phase}) gives the rough location of the Majorana mode. In Fig. \ref{fig:fig5}, we plot this location as a function of $t$, for fixed $\Delta=50E_r$, $\Omega=E_r$, $\mu_{\rm SF}=50E_r$, $\delta=49E_r$, and the harmonic trap $V(x)=m\omega^2x^2/2$ with $\hbar\omega=0.1E_r$. We can see that the system is always topologically trivial at the edge. The center is trivial at large $t$ and non-trivial at small $t$. Typically one goes from non-trivial to trivial as one moves from the center to the edge. For these parameters, the entire cloud is topologically trivial when $t>t_c\approx 8.4E_r$. We also plot the locations and widths of the Majorana modes from our numerical simulations in Fig. \ref{fig:fig3}. We can see that this simple argument gives a good approximation to the location of the Majorana modes, even for the small system we are considering.

\section{Heating}

One reason for introducing this model was to control heating. Only the Raman-coupled atoms experience heating from off-resonant light scattering. The energy absorbed from one scattering event is roughly the recoil energy $E_r\approx\hbar\times 50{\rm kHz}$. The number of scattering events per unit time is $N_{\rm R}\times\Gamma_s$, where the single particle inelastic scattering rate is $\Gamma_s$. As shown by Wei and Mueller, the inelastic scattering rate is proportional to the Raman coupling strength \cite{Wei2}, and as seen in Fig. \ref{fig:fig2}, for our states $\Gamma_s\approx\Omega/400$ . A typical Raman strength is $\Omega\sim E_r$, yielding an absorbed power of $P=N_{\rm R}E_r\Gamma_s\approx 125N_{\rm R}E_r/{\rm s}$. Due to atomic collisions this energy gets redistributed among all the atoms. Superfluidity will be destroyed when the energy per atom $\Delta E=Pt/(N_{\rm S}+N_{\rm R})$ is of order $k_BT_c\approx 0.3\mu_{\rm SF}$. Taking $\mu_{\rm RM}=E_r$, $\mu_{\rm SF}=50E_r$, and the harmonic potential $\hbar\omega=0.1E_r$, we have $N_{\rm S}\approx 500$, $N_{\rm R}\approx 20$, yields $N_{\rm R}/(N_{\rm S}+N_{\rm R})\approx 0.04$, and the system should remain superfluid for a time of order $t\approx 3{\rm s}$. By contrast, if all atoms experienced the Raman coupling, the lifetime would be $t\approx 120{\rm ms}$.

\section{Two- and three-body losses}
\begin{figure}[th]
  \includegraphics[width=0.45\textwidth]{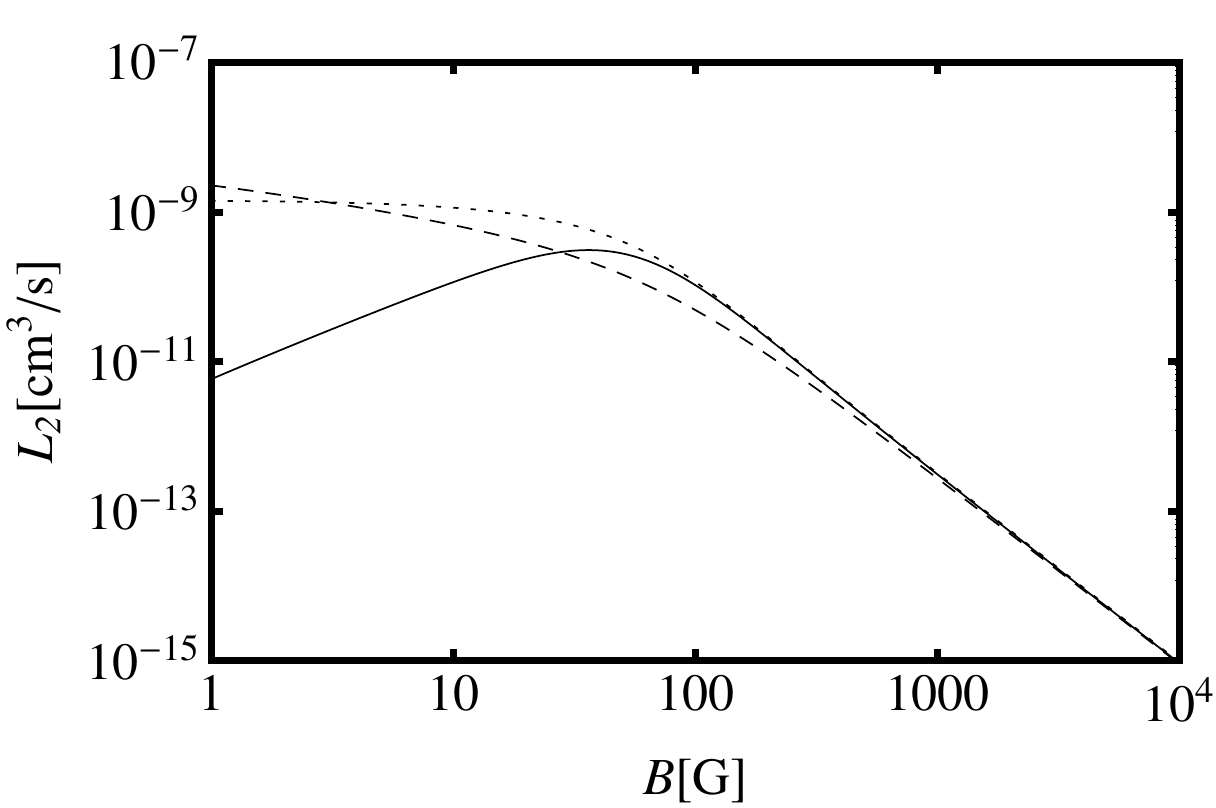}
  \caption{Exchange loss rate $L_2$ as a function of magnetic field $B$ using the formalism from \cite{Houbiers, OHara}. The four states are labeled 1 to 4 from the lowest to the forth lowest energy state. The solid line is collision from state $41\to21$, the dashed one is $42\to31$, and the dotted one is $43\to32$.}
  \label{fig:fig6}
\end{figure}

We must also consider inelastic processes coming from atomic collisions. For example, an atom in state 4 would collide with one in state 1, flipping into state 2. This process is allowed because it conserves $m_F$. The energy from such a spin-flip is sufficiently large that the atom would be lost from the trap. Considering our states, there are three such allowed exothermic processes: $41\to21$, $42\to31$, and $43\to 32$. The rate of atom loss is parameterized by the two-body coefficients $L_2$, appearing in the rate equation
\begin{eqnarray}
\dot{N}=-L_2N\langle n\rangle,
\end{eqnarray}
where $\langle n\rangle$ is the average density, and $N$ is the particle number. In Fig. \ref{fig:fig6} we plot the exchange loss rate coefficient $L_2$ as a function of magnetic field using asymptotic boundary condition approximation \cite{Houbiers, OHara}. Near the Feshbach region $B\approx832{\rm G}$, the two-body loss rate coefficient goes to $L_2\approx5\times10^{-13}{\rm cm^3/s}$. Thus for a system with density $n\sim 10^{11}{\rm cm^{-3}}$, the $1/e$ lifetime is about $20{\rm s}$.

We must also consider three-body losses, which are enhanced by the large mutual interaction between the lowest three hyperfine states near Feahbach resonance at $B\approx832{\rm G}$. Experiment observations show that $\dot{N}=-L_3N\langle n^2\rangle$ with the three-body loss rate coefficient is $L_3\approx 5\times10^{-22}{\rm cm^6/s}$ \cite{Huckans}. Thus for a low density system with density $n\sim10^{11}{\rm cm^{-3}}$, the $1/e$ lifetime is about $0.2{\rm s}$.

\section{summary}
In summary, we have devised an improved protocol to produce topological edge modes in Raman-induced spin-orbit coupled fermions. We divide the atoms into two populations, distinguished by their hyperfine states, yet spatially overlapping. The majority of the atoms experience strong interactions, while a minority experiences a Raman-induced spin-orbit coupling. Radio and microwaves mix the various states. By calculating the spatially resolved single particle density of states, we showed that this system has Majorana edge modes. We estimated heating rates from off-resonant light scattering, finding superfluid lifetimes of order of $3{\rm s}$ for $^6{\rm Li}$. We also estimate the two and three-body loss rate in the system, finding that the timescale is limited by the three-body losses: for a density of $n\sim10^{11}{\rm cm^{-3}}$, the $1/e$ lifetime is $0.2{\rm s}$.

The Majorana modes may be detected spectroscopically, with the caveat that the signal from a single mode will be weak. This can be improved by working with an array of quasi-1D tubes \cite{Liao}.

\begin{acknowledgements} 
We acknowledge useful comments from M. W. Zwierlein. J.X. would like to thank R. Wei for helpful discussions and providing the data shown in Fig. \ref{fig:fig2}. This research is supported by the National Science Foundation (PHY-1068165), the National Key Basic Research Program of China (Grant No. 2013CB922000), the National Natural Science Foundation of China (Grant No. 11074021), and the Army Research Office with funds from the DARPA OLE program. J.X. is also supported by China Scholarship Council.
\end{acknowledgements}

\end{document}